\documentclass[twocolumn, prl, showpacs, superscriptaddress]{revtex4}
\usepackage{amsfonts, amssymb, amsmath, graphicx, comment, bm, slashed, 
dcolumn, color}
\newcommand \la{\raisebox{-.5ex}{$\stackrel{<}{\sim}$}}

\newcommand \beq{\begin{eqnarray}}
\newcommand \eeq{\end{eqnarray}}
\newcommand \be{\begin{equation}}
\newcommand \ee{\end{equation}}

\def\simge{\mathrel{%
       \rlap{\raise 0.511ex \hbox{$>$}}{\lower 0.511ex \hbox{$\sim$}}}}
\def\simle{\mathrel{
       \rlap{\raise 0.511ex \hbox{$<$}}{\lower 0.511ex \hbox{$\sim$}}}}

\begin{document}
\title{Superfluid transition temperature of spin-orbit and 
\\ Rabi coupled fermions with tunable interactions}

\author{Philip D. Powell}
\affiliation{Lawrence Livermore National Laboratory, 7000 East Ave., 
Livermore, CA 94550, USA}
\affiliation{Department of Physics, University of Illinois at Urbana-Champaign, 
1110 W. Green Street, Urbana, Illinois 61801, USA}

\author{Gordon Baym}
\affiliation{Department of Physics, University of Illinois at Urbana-Champaign, 
1110 W. Green Street, Urbana, Illinois 61801, USA}

\author{C. A. R. S\'{a} de Melo}
\affiliation{School of Physics, Georgia Institute of Technology, 
837 State Street, Atlanta, Georgia 30332, USA}

\date{\today}

\begin{abstract} 
We obtain the superfluid transition temperature of equal Rashba-Dresselhaus spin-orbit and Rabi coupled 
Fermi superfluids, from the Bardeen-Cooper-Schrieffer (BCS) 
to Bose-Einstein condensate (BEC) regimes in three dimensions. 
Spin-orbit coupling enhances the critical temperature in 
the BEC limit, and can convert a first order 
phase transition in the presence of Rabi coupling into second order, as a function of the Rabi coupling
for fixed interactions.  We derive the Ginzburg-Landau equation to sixth power 
in the superfluid order parameter to describe both first and second order transitions as a function of spin-orbit and Rabi
couplings.
\end{abstract}

\pacs{67.85.Lm, 03.75.Ss, 47.37.+q, 74.25.Uv, 75.30.Kz}

\maketitle

The ability to simulate magnetic fields in cold atoms systems opens the possibility of
exploring new physics unachievable elsewhere.  In addition to artificial
Abelian magnetic fields~\cite{Lin2009,Lin2009-2,Ketterle2015}, one can also generate non-Abelian 
fields in both bosonic and fermionic systems~\cite{Lin2011,zwierlein2012,wang2012,Spielman2013,Galitski2013,Zhang2014,fallani2015,huang2016,wu2016}.   
The latter will eventually lead to the possibility of 
simulating quantum chromodynamics lattice gauge theory~\cite{cirac2012,Wiese-2013,Zohar2015,Dalmonte2016}. 
Present experiments on three dimensional spin-orbit coupled Fermi gases are still at 
too high a temperature for these systems to undergo Bardeen-Cooper-Schrieffer (BCS) pairing, because the current Raman scheme causes heating.
In contrast, theory has concentrated at zero 
temperature~\cite{Zhang2011,Zhai2011,Pu2011,Han2012,
Seo2012a,Seo2012b}.     
Once  such fermionic systems can be cooled below the superfluid transition temperature, the spin-orbit
coupling is expected to reveal new states with non-conventional pairing.   Even weak spin orbit coupling will produce an admixture of s-wave and p-wave pairing. 

In this Letter, we investigate the transition temperature 
of Fermi superfluids with an equal mixture of Rashba and Dresselhaus  spin-orbit coupling as a function
of the Rabi coupling, throughout the 
entire BCS (Bardeen-Cooper-Schrieffer)-to-BEC (Bose-Einstein condensation) evolution in three dimensions; 
the single particle Hamiltonian matrix is
\begin{equation}
   {\bf H}_{so} (\hat{\bf p}) =  \frac{(\hat{p}_x-\kappa \sigma_y)^2}{2m}  + \frac{\hat{p}_y^2}{2m} +  \frac{\hat{p}_z^2}{2m} -\frac{\Omega}{2}\sigma_z;
\end{equation}
the Pauli sigma matrices operate in the two-level space, 
$\hat{\bf p}$ is the momentum, $\Omega$ is the Rabi frequency,
and $\kappa$ is the momentum transfer to the atoms in a two-photon Raman process~\cite{Spielman2013}.
 This problem bears a close relation to spin-orbit coupling in solids, where the coupling $\sim p_i\sigma_j$  
is intrinsic, and where the role of the Rabi frequency is played by an external Zeeman magnetic field. 
While a mean field treatment describes well the evolution from the BCS to the BEC regime at zero 
temperature~\cite{Leggett1980}, this order of approximation fails to describe the correct critical temperature of the system 
in the BEC regime, because the physics of two-body bound states 
(Feshbach molecules)~\cite{Carlos1993} is not captured when the pairing order parameter goes to zero.   To remedy this problem,  we
include effects of order-parameter fluctuations in the thermodynamic potential.

 We stress that our present results are applicable to both neutral cold atomic and charged condensed matter systems.   We find that the spin-orbit coupling can enhance the critical temperature of 
the superfluid in the BEC regime and that it can convert a 
discontinuous first order phase transition in the presence of Rabi coupling into a continuous 
second order transition, as a function of the Rabi frequency (or Zeeman field in solids) for fixed interactions. 
We analyze the nature of the phase transition in terms of the Ginzburg-Landau free energy, calculating it to six powers 
of the superfluid order parameter to allow for the description 
of continuous and discontinuous transitions as a function of the spin-orbit 
coupling, Rabi frequency, and interactions.

To describe three dimensional Fermi superfluids in the presence of 
spin-orbit and Zeeman fields, we start from the Hamiltonian density
\begin{equation}
\label{eqn:hamiltonian}
{\cal H} ({\bf r}) = {\cal H}_{so} ({\bf r}) + {\cal H}_I ({\bf r}), 
\end{equation}
and use units $\hbar = k_B = 1$. 
The first term in Eq.~(\ref{eqn:hamiltonian}) is
the independent-particle contribution including spin-orbit coupling,
\begin{equation}
{\cal H}_{so} ({\bf r}) = \sum_{s s^\prime} \psi^\dagger_s ({\bf r}) 
\left[{\bf H}_{so}({\hat {\bf p}})\right]_{s s^\prime} \psi_{s^\prime} ({\bf r}),
\end{equation}
The second term describes the two-body s-wave contact interaction  
\begin{equation}
\label{eqn:interaction-hamiltonian}
{\cal H}_I ({\bf r}) 
= 
- 
g 
\psi^\dagger_{\uparrow} ({\bf r}) 
\psi^\dagger_{\downarrow} ({\bf r}) 
\psi_{\downarrow} ({\bf r}) 
\psi_{\uparrow} ({\bf r}),
\end{equation}
where the arrows indicate the pseudospins of the fermions, 
which we refer to simply as ``spins".
Here $g > 0$ corresponds to a constant attraction between opposite spins.  

  The pairing field 
$
\Delta ({\bf r},\tau) 
= 
- 
g 
\langle 
\psi_\downarrow ({\bf r},\tau) 
\psi_\uparrow ({\bf r},\tau)
\rangle
$ 
describes the formation of pairs of two fermions with opposite \
spins, where $\tau = it$ is the imaginary time.  Standard manipulations lead to the Lagrangian density
\begin{eqnarray}
{\cal L} ({\bf r, \tau}) & = & 
\frac{1}{2} 
\hspace{.5mm} 
\Psi^\dagger ({\bf r},\tau) 
{\bf G}^{-1} ({\bf r},\tau) 
\Psi ({\bf r},\tau) 
+ \frac{|\Delta ({\bf r},\tau)|^2}{g}   
\nonumber   
\\
&& \hspace{5mm} 
+ 
K ({\bf r}) \delta ({\bf r} - {\bf r}^\prime),   
\label{eq:Leff}
\end{eqnarray}
where 
$
\Psi 
= 
(\psi_{\uparrow} 
\hspace{1mm} 
\psi_{\downarrow} 
\hspace{1mm} 
\psi^\dagger_{\uparrow} 
\hspace{1mm} 
\psi^\dagger_{\downarrow})^T
$
is the Nambu spinor, 
and %
$
K \equiv
-\nabla^2 /2 m 
- 
\mu
$ 
is the kinetic energy operator measured with respect to the fermion
chemical potential $\mu$. We note that the definition of $\mu$
already includes the overall positive shift 
$\kappa^2/2m$ in the single particle kinetic energies 
due to spin-orbit coupling, that is, $\mu$ is 
measured with respect to $\kappa^2/2m$.

The inverse Green's function appearing in Eq.~(\ref{eq:Leff}) is
\begin{eqnarray}
{\bf G}_{\bf k}^{-1} (\tau) 
= 
\begin{pmatrix}			
\partial_\tau - K_\uparrow &- i\kappa k_x/m & 0 & -\Delta \\
i\kappa k_x/m & \partial_\tau - K_\downarrow & \Delta & 0 \\
0 & \Delta^* & \partial_\tau + K_\uparrow & -i\kappa k_x/m \\
-\Delta^* & 0 & i\kappa k_x/m & \partial_\tau + K_\downarrow
\end{pmatrix},   \nonumber\\
\label{eq:GF}
\end{eqnarray}
where 
$ K_{\uparrow} = K -\Omega/2,
$ 
and 
$
K_{\downarrow} = K + \Omega/2
$  
are the kinetic energy terms shifted by the Rabi coupling.  As mentioned above, the mean field treatment fails to describe the correct critical temperature of the system 
in the BEC regime.   
To incorporate the physics of two-body bound states, 
we must include effects of order-parameter fluctuations in the thermodynamic potential.

To obtain the transition temperature to the superfluid state, 
we analyze the partition function 
$
{\cal Z}$ as a functional integral $ 
\int {\cal D}\Delta {\cal D}\Delta^*
\int {\cal D}\Psi {\cal D}\Psi^\dagger e^{\cal S}$
for the Fermi superfluid, where
${\cal S} = \int_0^\beta \int d^3 {\bf r} {\cal L}({\bf r}, \tau)$
is the full action of the system. Upon integration over the fermion fields, 
the thermodynamic potential $\Omega = -T \ln {\cal Z}$ contains two terms
$\Omega = \Omega_0 + \Omega_F$, where
$\Omega_0 = -T \ln {\cal Z}_0 = -T S_0$ is the saddle point 
contribution, at which point $\Delta ({\bf r}, \tau) = \Delta_0$, 
and $\Omega_F = - T \ln {\cal Z}_F$ is the 
fluctuation part.  
The subscript $0$ denotes quantities calculated in mean field.

The mean-field (saddle-point) term in the thermodynamic potential is
\begin{equation}
\Omega_0 
= 
V\frac{|\Delta_0|^2}{g} 
- 
\frac{T}{2} \sum_{{\bf k},j} 
\ln
\left[
1 + e^{-\beta E_{j} ({\bf k})} 
\right]
+ 
\sum_{\bf k} \xi_{k},
\end{equation}
where $\xi_{k} = \varepsilon_{k}  - \mu$, $\varepsilon_{k} = {k}^2 /2 m$,
and the $E_{j} ({\bf k})$ are the eigenvalues of the 
Nambu Hamiltonian matrix 
${\bf H}_0 ({\bf k}) = {\bf 1} \partial_\tau - {\bf G}_{\bf k}^{-1} (\tau)$, 
with $j = \{1,2,3,4\}$. The first set of eigenvalues 
\begin{equation}
E_{1,2} ({\bf k})
= \sqrt{E_{0,k}^2 +  h_{\bf k}^2\pm 2 \sqrt{E_{0,k}^2  +  h_{\bf k}^2 - 
|\Delta_0|^2 (\kappa k_x/m)^2}}, 
%\nonumber\\ 
\label{ev}
\end{equation}
describe quasiparticle excitations, and the second set of eigenvalues 
$
E_{3,4} ({\bf k})
= 
- 
E_{1,2} ({\bf k})
$
correspond to quasiholes.
Here
$
E_{0,k} 
= \sqrt{
\xi^2_{k} 
+ 
|\Delta_0|^2},
$
and $ h_{\bf k} \equiv \sqrt{ (\kappa k_x/m)^2
 + \Omega^2/4}$ is the magnitude of the combined spin-orbit and
Rabi couplings.

The order parameter equation is found from the saddle point condition
$\delta \Omega_0 / \delta \Delta_0^* \vert_{T, V, \mu} = 0$, leading to
\begin{equation}
\frac{m}{4 \pi a_s} 
= 
\frac{1}{2 V} 
\sum_{\bf k} 
\bigg[ 
\frac{1}{\varepsilon_{k}}
- 
A_{+}
- 
\frac{h^2_z}{\xi_{k} h_{\bf k}} 
\hspace{.5mm} 
A_{-} 
\bigg].   
\label{eq:op_eqn}
\end{equation}
Here, we  write the interaction $g$ in terms of the renormalized
$s$-wave scattering length $a_s$ via the relation 
$1/g = - m/ 4 \pi a_s + (1/V)\sum_{\bf k} 1 / 2\varepsilon_{k}$
~\cite{as_note, Goldbart2011, Ozawa2012}, and for short write $
A_{\pm} 
= (1 - 2 n_{{\bf k},1}) / 2 E_1 
\pm 
 (1 - 2 n_{{\bf k},2}) / 2 E_2, 
$ with 
$n_{{\bf k},i} 
= 
1 / (e^{\beta E_{{\bf k},i}} + 1)
$
the Fermi function.   In addition, the particle number at the saddle point $
N_0 
= 
- \partial \Omega_0 / \partial \mu \vert_{T, V},
$ 
is given by
\begin{equation}
N_0 
=  
\sum_{\bf k} 
\bigg \{ 
1 - \xi_{k}
\bigg[ 
A_+ 
+ 
\frac{(\kappa k_x/m)^2}{\xi_{k} h_{\bf k}} 
\hspace{.5mm} 
A_- 
\bigg]  
\bigg \}.   
\label{eq:num_eqn}
\end{equation}

  The saddle point transition temperature $T_0$ is 
determined by solving Eq.~(\ref{eq:op_eqn}) for given $\mu$.     The corresponding number
of particles is given by Eq.~(\ref{eq:num_eqn}). 
This mean field treatment leads to a transition temperature growing as $e^{1/k_Fa_s}$ for $k_Fa_s \to 0^+$.  To find the physically correct transition temperature we must, in constructing the thermodynamic potential, include the physics of two-body bound states 
near the transition via the two-particle t-matrix
~\cite{Nozieres1985,Baym2006}.
With all the two particle channels taken into account, 
the t-matrix calculation leads to a two-particle scattering amplitude, 
$\Gamma$, where
\begin{equation}
\Gamma^{-1} ({\bf q},z) 
= 
\frac{m}{4 \pi a_s} 
- 
\frac{1}{2V} \sum_{\bf k} 
\bigg[ 
\frac{1}{~\varepsilon_{k}} 
+ 
\sum^2_{i,j=1} 
\alpha_{ij} W_{ij} 
\bigg];  
\label{eq:Gamma}
\end{equation}
here $z$ is the (complex) frequency, 
$
W_{ij} 
= 
(1 - n_{{\bf k},i} - n_{{\bf k+q},j}) 
/ 
(z - E_i({\bf k}) - E_j({\bf k +q}) ).
$ 
In the limit that  
the order parameter goes to 0, the single particle eigenvalues reduce to 
$E_{1,2}({\bf k}) = \xi_k \pm h_{\bf k}$~\cite{eigenvalue_note}.  
The coefficients
$
\alpha_{11} 
= 
\alpha_{22} 
= 
\vert
u_{\bf k} u_{{\bf k}+{\bf q}} 
- 
v_{\bf k} v^*_{{\bf k}+{\bf q}}
\vert^2
$
and 
$
\alpha_{12} 
= 
\alpha_{21} 
= 
\vert 
u_{\bf k} v_{{\bf k}+{\bf q}} 
+ 
u_{{\bf k}+{\bf q}} v_{\bf k}
\vert^2
$ 
are weighting functions of the amplitudes
\begin{equation}
u_{\bf k} 
= 
\sqrt{\frac{1}{2} 
\left(
1 + \frac{\Omega}{2h_{\bf k}} 
\right)},   
\hspace{5mm}   
v_{\bf k} 
= i\sqrt{\frac{1}{2} 
\left(1 - \frac{\Omega}{2h_{\bf k}} 
\right)}.
\end{equation}

  As the fermion chemical potential becomes large and negative, the system becomes 
non-degenerate and $\Gamma^{-1}({\bf q}, z) = 0$ 
becomes the exact eigenvalue equation for the two-body bound state in the presence of spin-orbit and Rabi coupling~\cite{Doga2016}. The solution is $z= E_{bs}({\bf q})  - 2 \mu$, where $E_{bs}(\bf q) $ is the  two-body bound state energy.  The fluctuation correction to the thermodynamic potential is then
$
\Omega_F 
= 
- 
T 
\sum_{{\bf q},i q_n} 
\ln 
\left[ 
\beta
\Gamma ({\bf q},i q_n)/V
\right].
$

  From $\Omega_F$ we obtain the fluctuation contribution to the particle number
$N_F = - \partial \Omega_F / \partial \mu \vert_{T, V} = N_{sc} + N_b$.
Here, 
\begin{equation}
N_{sc} 
=  
\sum_{\bf q} 
\int^\infty_{\omega_{tp} ({\bf q})} 
\frac{d \omega}{\pi} 
n_B (\omega) 
\left[
\frac{\partial \delta ({\bf q},\omega)}{\partial \mu}
-\frac{\partial \delta ({\bf q},0)}{\partial \mu}
\right]_{V,T}  
\label{eq:nscattering}
\end{equation}
is the number of particles in scattering states,
where the phase shift $\delta ({\bf q},\omega)$ is
defined via the relation 
$
\Gamma ({\bf q},\omega \pm i \epsilon) 
= 
|\Gamma ({\bf q},\omega)| e^{\pm i \delta({\bf q},\omega)}
$ 
and $\omega_{tp} ({\bf q})$ is the two-particle continuum threshold 
corresponding to the branch point 
of $\Gamma^{-1} ({\bf q},z)$~\cite{Nozieres1985, Pethick2011}.
Also, 
\begin{equation}
N_b  
=
2 
\sum_{\bf q} n_B (E_{bs} ({\bf q})),   
\label{eq:nbound}
\end{equation}
is the number of fermions in bound states  with
$n_B (\omega) = 1 / (e^{\beta \omega} - 1)$ 
the Bose distribution function. The total number of fermions, as a function of $\mu$, becomes 
\begin{equation}
\label{eq:number-total}
N = N_0 + N_F,
\end{equation}
where $N_0$ is
given in Eq.~(\ref{eq:num_eqn}) and $N_F$ is the 
sum the two contributions $N_{sc}$ and $N_b$ 
discussed above~\cite{Nozieres1985,Carlos1993}.%
\begin{figure}[htb]
\includegraphics[width=0.45\textwidth,height=0.3\textwidth]{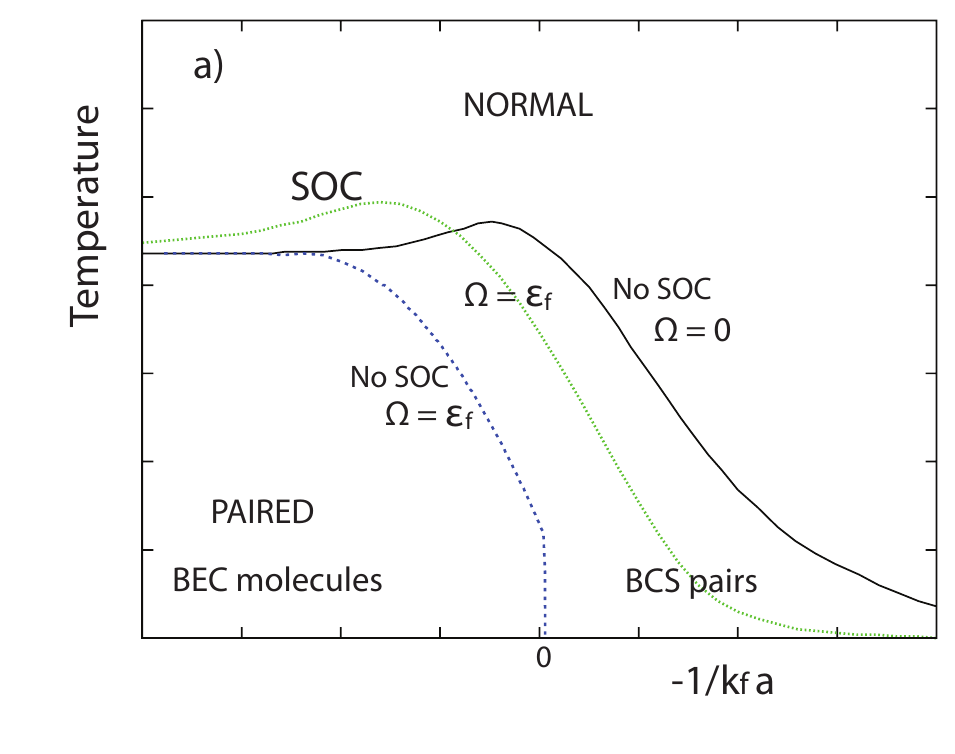}
\includegraphics[width = 0.45\textwidth,height=0.3\textwidth]{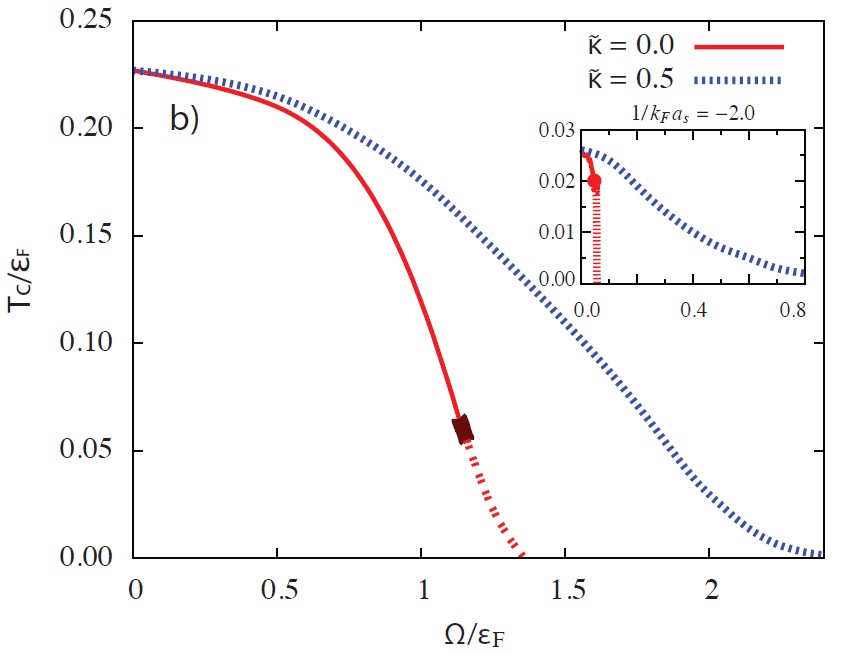}
\caption{
\footnotesize 
(Color online)
a) The transition temperature $T_c$ for ERD spin-orbit coupling
for two Rabi coupling strengths,  $\Omega = 0$ and $\varepsilon_F$.   
For $\Omega=0$,  solid (black) curve, $T_c$ is that for zero spin-orbit coupling, since the ERD field can be gauged away.
The dashed (blue) line shows $T_c$ for zero spin-orbit coupling, with $\Omega = \varepsilon_F$,  while the dotted (green) line shows $T_c$ for $\Omega = 0$ and $\varepsilon_F$,   
and $\kappa = 0.5 k_F $.
In b) $T_c$ is drawn at unitarity, 
$1/k_F a_s = 0$, and in the inset at $1/k_F a_s = -2.0$, 
as a function of $\widetilde\Omega = \Omega/\varepsilon_F$.  
The solid (red) curves are for
$\tilde\kappa = 0$, and the dashed (blue) curves are 
for $\tilde\kappa = 0.5$.  Across the dotted (red) curves below the solid (red) curves, the phase transition is 
first order.   
}
\label{fig:figure1}
\end{figure}

Defining $k_F$ to be the Fermi momentum of the
atomic gas with total density $n = k_F^3/3\pi^2$, 
we obtain $T_c$ as a function of the scattering parameter $1/k_Fa_s$ 
by solving simultaneously the order parameter
and number equations (\ref{eq:op_eqn}) 
and~(\ref{eq:number-total}). 
Figure~\ref{fig:figure1} shows the effects of spin-orbit 
and Rabi couplings on the transition temperature $T_c$.  
The solutions correspond to minima of the free energy 
${\cal F} = \Omega + \mu N$.   
In Fig.~\ref{fig:figure1}, 
we scale energies and temperatures by the Fermi energy 
$\varepsilon_F = k_F^2/2m$.
   
The solid (black) line in Fig.~\ref{fig:figure1}a  
shows the transition temperature $T_c$ between the normal
and superfluid state versus the
scattering parameter $1/k_F a_s$ for zero Rabi coupling $(\Omega = 0)$
and zero one-dimensional 
Rashba-Dresselhaus (ERD)~\cite{Dresselhaus1955, Rashba1960, footnote1} 
spin-orbit coupling $(\kappa = 0)$.
If $\Omega = 0$, the spin-orbit coupling $\kappa$
can be removed by a simple gauge transformation, and thus plays no role.   In this situation,
the pairing is purely s-wave.
The dashed (blue) line shows $T_c$ for $\Omega \ne 0$, with vanishing
ERD spin-orbit coupling.   We see that for fixed interaction strength, the pair-breaking effect 
of the Rabi coupling (as a Zeeman field breaks pairs in a
superconductor) suppresses superfluidity, compared with $\Omega = 0$.
With both ERD spin-orbit and Rabi coupling present, the pairing is no longer pure s-wave, but has a
triplet p-wave component (and higher) mixed into the
superfluid order parameter; the admixture stabilizes the superfluid phase, as
shown by the dotted (green) line. 
The latter curve shows that  in the BEC regime with large positive
$1/k_F a_s$, the transition temperature $T_c$ is larger with spin-orbit and Rabi couplings than in their absence, as a consequence of the reduction 
of the bosonic effective mass in the x-direction below $2m$.
However, with sufficiently large $\Omega$, the geometric mean bosonic mass 
$M_B$ increases and 
$T_c$ decreases~\cite{footnote:mass-renormalization}.

Figure~\ref{fig:figure1}b shows $T_c$ versus $\Omega$ 
for fixed $1/k_Fa_s$, without and with
ERD spin-orbit coupling at $\kappa = 0.5 k_F$.
When $\kappa$ and the temperature are zero, superfluidity is destroyed 
at a critical value of $\Omega$ corresponding to the 
Clogston limit~\cite{clogston1962}.  
At low temperature the phase transition to the normal state is first order, 
because the Rabi coupling (Zeeman field) is sufficiently large to 
break singlet Cooper pairs. 
However, at higher temperatures the singlet s-wave superfluid starts 
to become polarized due to thermally excited quasiparticles that produce 
a paramagnetic response.   Therefore above the characteristic temperature 
indicated by the large (red) dots, the transition becomes second order,
as pointed out by Sarma~\cite{sarma1963}. The critical temperature for $\kappa \ne0$ vanishes only asymptotically 
in the limit of large $\Omega$.   
We note that for $\Omega = E_F$ and $\kappa = 0$ the transition from the superfluid
to the normal state is continuous at unitarity, but very close to a discontinuous
transition.  In the range $1.05 \lesssim \Omega/E_F \lesssim 1.10$
numerical uncertainties as $\kappa\to 0$ prevent 
us from predicting exactly whether the transition 
at unitarity is continuous or discontinuous.

To understand further the effects of fluctuations on the 
order of the transition to the superfluid phase and to  
assess the impact of spin-orbit and Rabi couplings near the 
critical temperature,
we now derive the 
Ginzburg-Landau description of the free energy near the transition, where
the action ${\cal S}_F$ can be expanded in powers of the 
order parameter $\Delta ({q}) $, beyond
Gaussian order.   The expansion of ${\cal S}_F$ to quartic power is sufficient to describe the continuous (second order) transition in $T_c$ versus $1/k_F a_s$ in the absence of an external Zeeman field~\cite{Carlos1993}.  However, to describe correctly the first order 
transition~\cite{clogston1962,sarma1963} at low temperature (Fig.~\ref{fig:figure1}), it is necessary to
expand the free energy to sixth order in $\Delta$.

The quadratic (Gaussian order) term in the
action is 
\begin{eqnarray}
 {\cal S}_G
& = & \beta V \sum_q \frac{|\Delta_q|^2}{\Gamma ({\bf q}, z)}.
\end{eqnarray}
For an order parameter varying slowly in space and  time, we may expand 
\begin{equation}
\Gamma^{-1} ({\bf q},z) 
= a  + c_i \frac{q^2_i}{2 m}  - d_0  z  + \cdots, 
\label{eq:Gamma_expansion}
\end{equation}
with the sum over $i=x,y,z$ implicit.
The full result, as a functional of $\Delta(r,\tau)$, has the form
\begin{eqnarray}
 {\cal S}_F
& = & \int_0^\beta d\tau \int d^3r \Big(d_0\Delta^* \frac{\partial}{\partial \tau}\Delta+a|\Delta|^2 \nonumber\\
&&+c_i\frac{ | \nabla_i \Delta|^2}{2 m} + 
\frac{b}2 |\Delta|^4 + 
\frac{f}3 |\Delta|^6 \Big).
\end{eqnarray} 
The full time-dependent Ginzburg-Landau action describes 
systems in and near equilibrium, e.g., with collective modes.  The imaginary part of $d_0$ measures the non-conservation of $|\Delta|^2$ in time.

We are interested here in systems at 
thermodynamic equilibrium where the order parameter is 
independent of time.  Then minimizing the free energy $T{\cal S}_F$ with respect to $\Delta^*$,
we obtain the Ginzburg-Landau equation 
\begin{equation}
\left(-\frac{c_i}{2 m}  \nabla^2_i + b |\Delta|^2 + f |\Delta|^4 + a \right) \Delta = 0 .   
\label{eq:TDGL}
\end{equation}
 For $b$ positive the system undergoes a continuous 
phase transition when $a$ changes sign.   
However, when $b$ is negative the system is unstable in the absence of $f$.  For $b<0$ and 
 $a > 0$, a first order phase transition
occurs when $3 b^2 = 16 a f$.
Positive $f$ stabilizes the system even when $b < 0$. 

In the BEC regime, we define an effective bosonic 
wavefunction $\Psi= \sqrt{d_0} \Delta$ to recast Eq.~(\ref{eq:TDGL}) 
in the form of the Gross-Pitaevskii equation for a dilute Bose gas 
\begin{equation}
\left( -
\frac{\nabla^2_i }{2 M_{i}} 
+ U_{2} |\Psi (r)|^2 
+U_{3} |\Psi (r)|^4 
-\mu_B \right)\Psi (r) = 0.   
\label{eq:GP}
\end{equation}
Here, $\mu_B = - a / d_0$ is the bosonic chemical potential, 
the $M_{i} = m (d_0 / c_i)$ are the anisotropic bosonic masses, 
and $U_{2} = b / d_0^2$ and $U_{3} = f/d_0^3$ represent contact
interactions of two and three bosons.  In the BEC regime these terms are   
always positive, thus leading to a system 
consisting of a dilute gas of stable bosons.   
The boson chemical potential $\mu_B$ is $\approx 2 \mu + E_b <  0$,
where $E_b$ is the two-body bound state energy in the presence
of spin-orbit coupling and Rabi frequency,  obtained from the condition 
$\Gamma^{-1} ({\bf q}, E - 2\mu)= 0$, 
discussed earlier. 

 The anisotropy of the effective bosonic masses, $M_{x} \ne M_{y} = M_{z} \equiv M_\perp$
stems from the anisotropy of the ERD spin-orbit coupling, which together 
with the Rabi coupling modifies the dispersion of the constituent 
fermions along the $x$ direction.  In the limit $k_Fa_s \ll 1$ the 
many-body effective masses reduce to those obtained by expanding the 
two-body binding energy 
$E_{bs}({\bf q}) \approx-E_b + q^2_i / 2 M_{i},$
and agree with known results~\cite{Doga2016}.
However, for $1/k_F a_s \,\la \, 2$, many-body and thermal effects produce 
deviations from the two-body result. 

In the absence of two and three-body boson-boson interactions ($U_2$ and $U_3$), we
directly obtain the analytic expression for $T_c$ in the 
Bose limit from Eq.~(\ref{eq:nbound}),
\begin{equation}
  T_c = \frac{2\pi}{M_B}\left(\frac{n_B}{\zeta(3/2)}\right)^{2/3}, 
\end{equation}
with $M_B= (M_xM_\perp^2)^{1/3}$,
 by noting that 
$\omega_p ({\bf q}) = E_{bs} ({\bf q})$
and using the condition that  $n_B \simeq n/2$  (with corrections exponentially small in $(1/k_Fa_s)^2$), where $n_B$ is the
density of bosons and $n$ is the density of fermions. 
In the BEC regime, the results shown in Fig.~\ref{fig:figure1}
include the effects of the mass anisotropy, but do not include effects 
of boson-boson interactions. 

To account for boson-boson interactions, we use the Hamiltonian of
 Eq.~(\ref{eq:GP}) with $U_2 \ne 0$, but with $U_3 = 0$, and 
 apply the method developed in Ref.~\cite{Baym1999} to show
that these interactions further increase $T_{BEC}$ to 
\begin{equation}
T_c (a_B) = (1 + \gamma) T_{BEC},
\end{equation}
where $\gamma = \lambda n_B^{1/3} a_B$. 
Here,  $a_{B}$ is the s-wave boson-boson scattering 
length,  $\lambda$ is a dimensionless constant $\sim 1$,
and we used the relation $U_2 = 4\pi a_B/M_B$. 
Since 
$n_B =  k_F^3/6 \pi^2$ 
and the boson-boson scattering length is
$a_B = U_2 M_B/4\pi$, we have
$\gamma = {\tilde \lambda} {\widetilde M}_B {\widetilde U}_2,$
where 
$
{\widetilde M}_B 
= 
M_B/2m,
$ 
$
{\widetilde U}_2 =
U_2 k_F^3/\varepsilon_F   ,
$
and 
$
{\tilde \lambda}
= 
\lambda
/
4(6\pi^5)^{1/3}
\approx 
\lambda/50 .
$
For fixed $1/k_F a_s$, $T_c$ is enhanced both by a spin-orbit and the $\Omega$ dependent 
decrease in the effective boson mass $M_B$ ($\sim$10-15\%), 
as well as a stabilizing boson-boson repulsion $U_2$ ($\sim$2-3\%),
for the parameters used in Fig.~\ref{fig:figure1}.

In summary, we have analysed the finite temperature phase diagram of 
three dimensional Fermi superfluids in the presence ERD spin-orbit coupling, 
Rabi coupling,  
and tunable s-wave interactions. Furthermore, 
we developed the Ginzburg-Landau theory up to sixth power 
in the amplitude of the order parameter to show the origin of  
discontinuous (first order) phase transitions when the Rabi frequency
is sufficiently large for vanishing spin-orbit coupling.

The research of author PDP was supported in part by  
NSF Grant PHY1305891 and that of GB by NSF Grants PHY1305891 and PHY1714042.  
Both GB and CARSdM thank the Aspen Center for Physics, 
supported by NSF Grants PHY1066292 and PHY1607611, where 
part of this work was done.  This work was performed under the auspices of the 
U.S. Department of Energy by Lawrence Livermore National Laboratory under 
contract DE- AC52- 07NA27344.


\begin{thebibliography}{99}
\bibitem{footnote:mass-renormalization}
This renormalization of the mass of the bosons can
be traced back to the actual change in the energy dispersion of the fermions
when both spin-orbit coupling and Zeeman fields are present.
\end{thebibliography}
\end{document}